\documentclass[letter]{aa} 

\usepackage{graphicx}
\usepackage{fixltx2e}
\usepackage{txfonts,booktabs}
\begin{document}
   \title{Candidate hypervelocity stars of spectral type G and K revisited}
   
   \titlerunning{Candidate Hypervelocity stars of spectral type G and K revisited}
   \authorrunning{Ziegerer et al.}

   \author{E. Ziegerer \inst{1} \and  M. Volkert \inst{1}\and U. Heber \inst{1} \and A. Irrgang \inst{1} \and B.T. G\"ansicke \inst{2} \and S. Geier \inst{3}$^,$\inst{1}}

   \institute{Dr.\,Remeis-Observatory \& ECAP, Astronomical Institute, Friedrich-Alexander University Erlangen-N\"urnberg, Sternwartstr.~7, 96049 Bamberg, Germany\\ \email{eva.ziegerer@sternwarte.uni-erlangen.de} \and Department of Physiks, University of Warwick, Coventry CV4 7AL, UK \and European Southern Observatory, Karl-Schwarzschild-Str. 2, 85748 Garching, Germany}

   \date{Received/ Accepted}

\abstract{
Hypervelocity stars (HVS) move so fast that they are unbound to the Galaxy. 
When they were first discovered in 2005, dynamical ejection from the supermassive black hole (SMBH) in the Galactic Centre (GC) was suggested as their origin. 
The two dozen HVSs known today are young massive B stars, mostly of 3--4 solar masses. 
Recently, 20 HVS candidates of low mass were discovered in the Segue G and K dwarf sample, but none of them originates from the GC. 
We embarked on a kinematic analysis of the Segue HVS candidate sample using the full 6D phase space information based on new proper motion measurements. 
Their orbital properties can then be derived by tracing back their trajectories in different mass models of our Galaxy. 
We present the results for 14 candidate HVSs, for which proper motion measurements were possible. 
Significantly lower proper motions than found in the previous study were derived. 
Considering three different Galactic mass models we find that all stars are bound to the Galaxy. 
We confirm that the stars do not originate from the GC. 
The distribution of their proper motions and radial velocities is consistent with predictions for runaway stars ejected from the Galactic disk by the binary supernova mechanism. 
However, their kinematics are also consistent with old disk membership. 
Moreover, most stars have rather low metallicities and strong $\alpha$-element enrichment as typical for thick disk and halo stars, whereas the metallicity of the three most metal-rich stars could possibly indicate that they are runaway stars from the thin disk. 
One star shows halo kinematics.
}
   \keywords{stars: kinematics and dynamics -- stars: low-mass -- stars: late-type -- stars: abundances -- stars: Population II }

   \maketitle
%

\section{Introduction}
\label{sec_introduction}
The high space velocities and large distances of hypervelocity stars (HVS) unbound to the Galaxy make them important probes to map the Galactic dark matter halo. 
When HVSs were first discovered \citep{Brown2005,Hirsch2005, Edelmann2005}, the tidal disruption of a binary by the supermassive black hole (SMBH) in the Galactic Centre (GC) was suggested as their origin \citep{Hills1988}. 
\citet{Brown2014} carried out a systematic survey for B-type stars in the halo and found about two dozen HVSs with intermediate masses in the range of 3 to 4 M$_\odot$.
As such stars are luminous their survey covered a large volume (out to 100\,kpc from the GC). 
Low-mass stars on the other hand can be accelerated more easily and may gain higher ejection velocities \citep{Tauris2014}. 
Since they are long-lived they can travel very large distances during their main-sequence lifetime. 
However, they are less luminous than the B-type HVSs and can only be detected in a smaller volume by flux-limited surveys ($<$10\,kpc), such as Sloan Extension for Galactic Understanding and Exploration (SEGUE).
Moreover, a photometric pre-selection of low mass main-sequence stars is very difficult because of the overwhelmingly large number of red stars in the halo. 
Therefore attempts have been made to isolate HVS candidates of low mass from the SEGUE  G and K Dwarf Sample  \citep[][hereafter P14]{Palladino2014}, LAMOST \citep{Zhong2014} and RAVE \citep{Hawkins2014} surveys using proper motion criteria. 

P14 carried out a search for G and K candidate-HVS from the SEGUE. 
High proper motion stars were selected for a detailed analysis of the 6D phase space information and 20 candidates were found likely to be unbound, four of which at 3$\sigma$ and six at 2$\sigma$ significance levels. 
Calculating the stars' trajectories in the Galactic potential P14 derived possible places of origin in the Galactic disk. 
Amongst the seven stars with the highest probability of being unbound ($>98\%$) none crossed the disk near the GC, but at distances of 5 to 10\,kpc away from it. 
Hence, an origin in the GC was excluded for those stars, challenging the SMBH slingshot mechanism. 
Other ejection mechanisms were discussed including the classical scenarii for dynamical interaction in star clusters and the binary supernova scenario \citep{Blaauw1961}.
The latter has been revisited by \citet{Tauris2014} to derive the maximum speed of HVS stars ejected from binaries. 
The simulations indicate that Galactic rest-frame velocities of up to 1280 kms$^{-1}$ are possible. 
Such high velocities can explain many, if not all, of the G/K-dwarf HVSs in the SEGUE sample.

As stressed by P14, the HVS nature of the stars depends strongly on the proper motion adopted. 
Therefore, the data was carefully checked for reliability by simulations. 
Three stars met all criteria and therefore were characterized as ``clean''.
The remaining 17 stars were regarded as ``reliable''. 
P14 found that the candidates' tangential velocities are much higher than their radial velocities unlike expected for an isotropic distribution of stars. 
The authors therefore caution that the high tangential- vs. radial-velocity ratio may be characteristic for a sample with large proper motion errors and built a Monte Carlo test to estimate the chance of the stars being outliers. 
All stars show a likelihood of less than 25\%, half of them even less than 10\%.
Nevertheless, an independent determination of the proper motions is required. 
 
In this letter we attempt to determine proper motions for all 20 stars of the sample by combining all astrometric information at hand from digitized plates and modern surveys.
We were able to derive proper motions for 14 stars of the sample.
Because of the strong implications of the results of P14 we also determined their radial velocities and distances and calculated their trajectories in different Galactic potentials. 
The confirmation of the candidates as HVSs would demand alternative ejection mechanisms other than the SMBH slingshot to explain their existence.


\section{Observations}
\label{sec_observations}

\subsection{Radial velocities and distances}
\label{sec_rv}

In order to reanalyse the spectra, we retrieved spectra of all 20 stars from the SDSS data base.  
In addition we obtained individual SDSS spectra to search for radial velocity variability in order to exclude close binaries. 
No statistically significant radial velocity variations were detected, i.e. no indication for binarity could be found. 
The averaged radial velocities were consistent with those given by P14 to within mutual error limits.

P14 made use of the DR9 SEGUE Stellar Parameter Pipeline (SSPP) which provided estimates for effective temperatures $T_\text{eff}$, surface gravity $\log g$, iron abundance [Fe/H] and $\alpha$ enrichment factor [$\alpha$/Fe]. 
They determined the distances of the stars using an isochrone-matching technique.
We checked effective temperatures and surface gravities for consistency by comparing $ugriz$-magnitudes from SDSS to synthetic colours from \citet{Castelli1999}.
Additionally we compared the synthetic spectra of \citet{Munari2005} to the SDSS spectra.
We found no inconsistencies and therefore adopted the atmospheric parameters and distances as given by P14.


\subsection{Proper motions}
\label{sec_pm}

The most problematic information for the kinematic analysis are the proper motions. 
P14 selected their sample using proper motions provided by the SDSS data base.

In order to determine the proper motions we made use of all available images including SDSS at different epochs.
Early epoch photographic plates from the Digitised Sky Surveys\footnote{http://archive.stsci.edu/cgi-bin/dss\_plate\_finder} were combined with those obtained from the data bases of modern digital surveys like SDSS\footnote{http://skyserver.sdss3.org/public/en/tools/chart/navi.aspx} and UKIDSS\footnote{http://www.ukidss.org/}.

\begin{figure}[t]
\begin{center}
\includegraphics[scale=0.31]{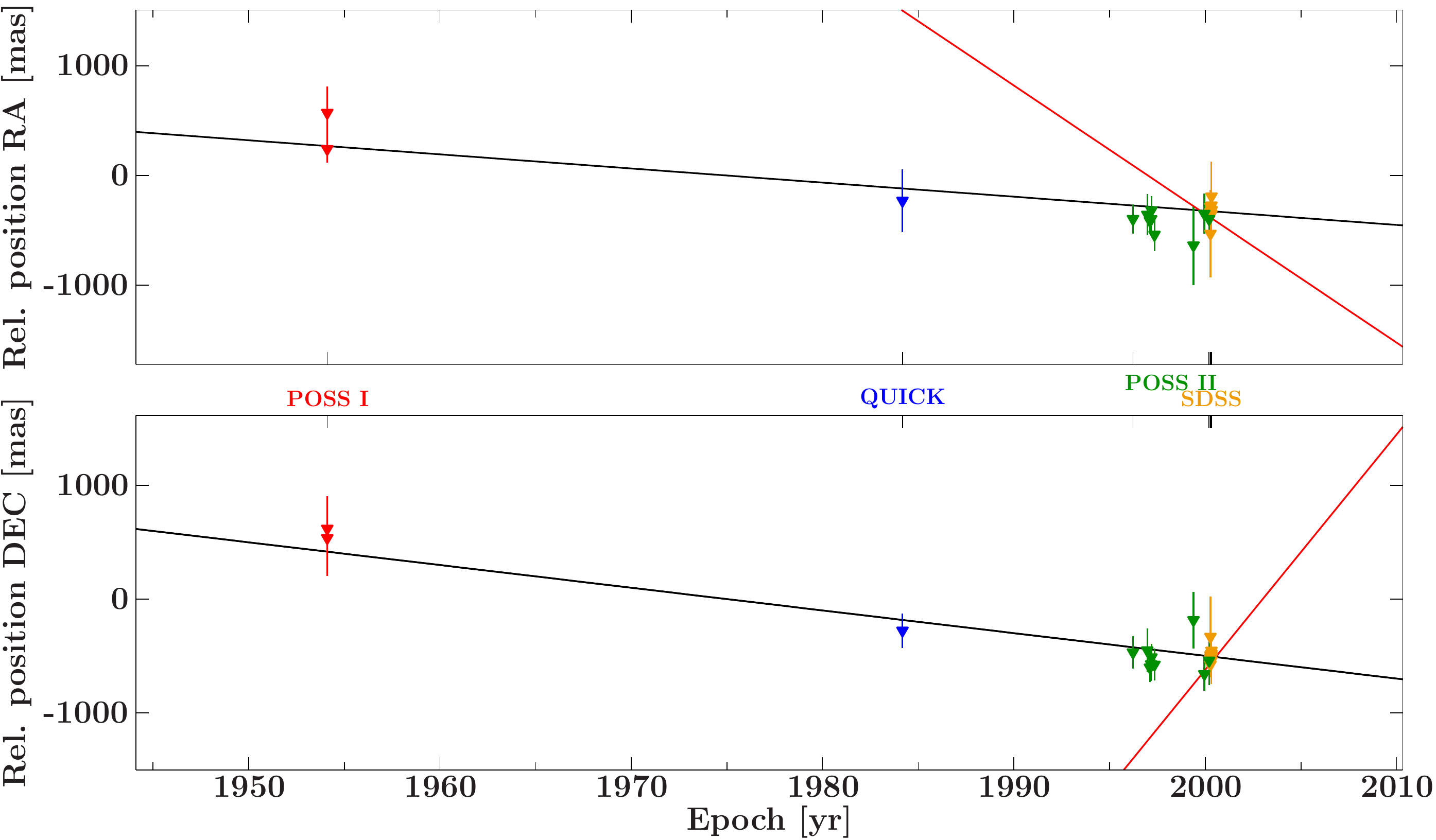}
\caption{\label{fig_15_PM}Proper motion fit through the relative positions of No.~4. Red lines indicate the proper motion used by P14.}
\end{center}
\end{figure}

For each HVS candidate, positions were derived from all available images with respect to a set of faint, compact and well distributed background galaxies (see \citealp{Tillich2011}).
This provides us with a timebase of about 60 years.
We were able to derive proper motions for 14 of the 20 P14 candidates.
For the low Galactic latitude stars No.~5, 7, 8, and 10 we failed to identify a sufficiently large set of suitable background galaxies. 
No. 1 and 11 suffered from crowding.
Therefore we had to dismiss these six stars from the further analysis. 
For the remaining candidates 16 to 29 background galaxies per field were found to be suitable and distributed favourably around the target. 
The proper motion components were obtained from the positions of all epochs by linear regression (see Fig. \ref{fig_15_PM}).
The values agree with the SDSS measurements for objects No.~2 and 18 only (see Table \ref{tab_pm}).

\begin{figure}[t]
\begin{center}
\includegraphics[scale=0.55]{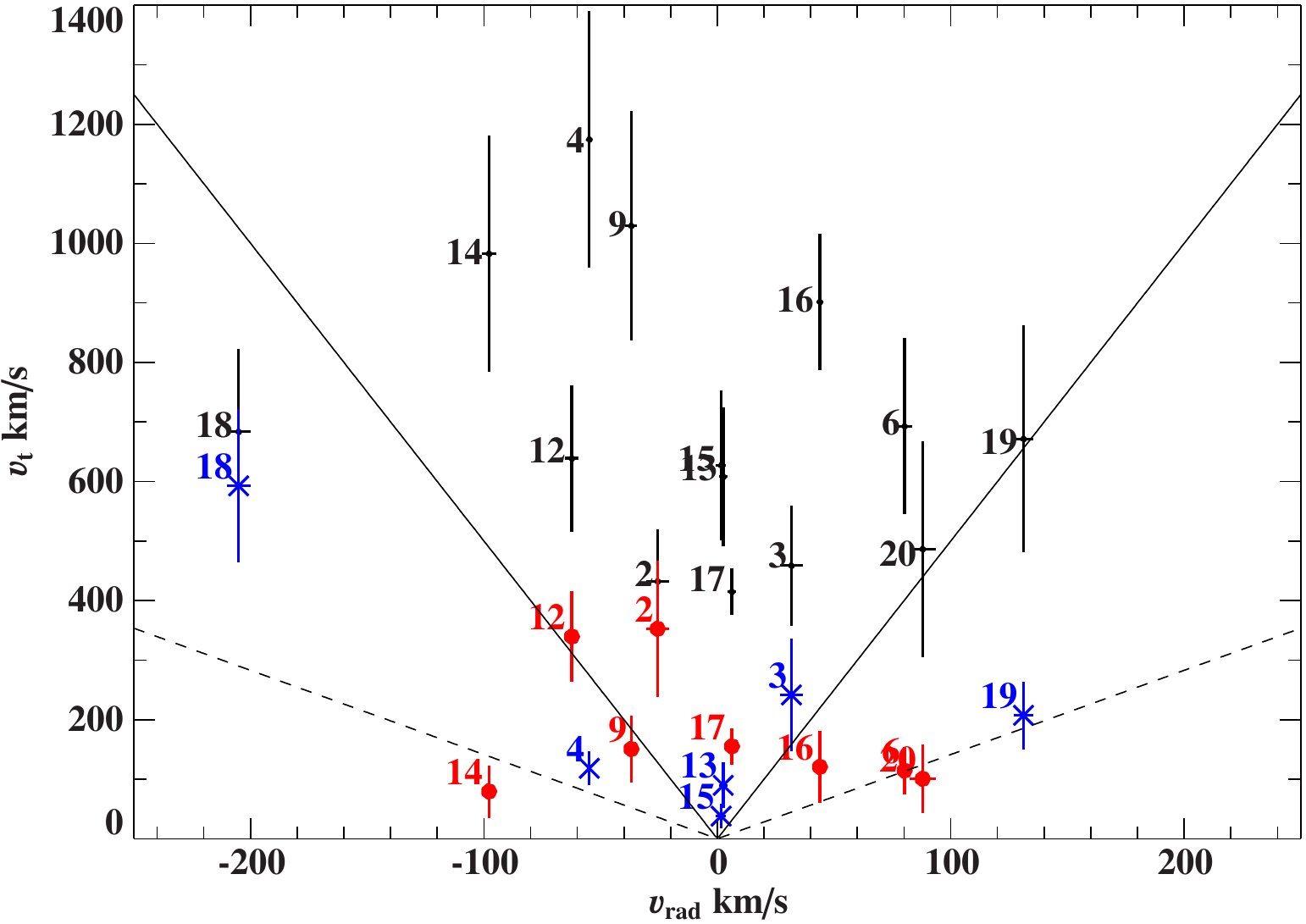}
\caption{\label{fig_vr_vt}Comparison of the tangential and radial velocity derived in this work (coloured with errorbars, red dots indicate stars with [Fe/H]$>-0.7$, blue crosses [Fe/H]$<-0.7$) and P14 (black). Error bars of P14 were derived in the same way as in this work. The dashed lines indicate $\varv_t=\sqrt2 \varv_r$, while the solid lines indicate $\varv_t=5 \varv_r$.}
\end{center}
\end{figure}

The tangential velocities are plotted in Fig. \ref{fig_vr_vt}  vs. the radial velocities and compared to the results of P14. 
Except for No.~2 and 18 the values derived here are significantly lower than those given by P14.
The ratios of tangential- vs. radial-velocity of the stars are much lower than previously found ($>$5).  
According to P14 an isotropic distribution of stars would be described by a $\sqrt2$ times higher tangential than radial velocity.
The new results come close to that expectation.

\begin{table*}
\scriptsize
\begin{tabular}{lcc|cc|ccccc|ccc}
 & \multicolumn{2}{c|}{this work} & \multicolumn{2}{c|}{SDSS} & \multicolumn{5}{c|}{} & \multicolumn{3}{c}{bound probabilities (\%)}\\
P14 & $\mu_\alpha\cos\delta$ & $\mu_\delta$ & $\mu_\alpha\cos\delta$ & $\mu_\delta$ & $\varv_\text{GRF,1}$ & $\varv_\text{GRF,2}$ & $z$ & $r$ & $r_\text{min}$ & P14 & \multicolumn{2}{c}{model II}\\
No. & (mas/yr) & (mas/yr) & (mas/yr) & (mas/yr) & (km/s) & (km/s) & (kpc) & (kpc) & (kpc) & \multicolumn{2}{c}{PM: SDSS} & this work \\
\hline\hline
2  & $8.9\pm4.7$   & $8.6\pm3.7$   & $-2.6\pm3.0$   & $15.5\pm3.0$  &$423.9\pm100.7$ & $635.2\pm86.7$&  $-2.9\pm0.2$ & $60.1$  & $3.4$  &  7.43   &  10.7 &   84.7  \\
3  & $5.5\pm4.0$   & $-10.4\pm4.7$ & $23.6\pm3.0$   & $-1.6\pm3.0$  &$303.7\pm75.9$ & $645.9\pm96.7$  & $2.5\pm0.5$  & $8.7$   & $1.5$   &  34.88  &  33.3 &   99.9  \\
4  & $-12.9\pm2.4$ & $-20.0\pm4.1$ & $-117.2\pm5.9$ & $206.8\pm5.9$ &$126.5\pm23.2$ & $1304.9\pm213.5$& $0.8\pm0.1$  & $6.7$   & $3.9$ &  0.00   &  0.0  &    100  \\
6  & $5.8\pm2.2$   & $-4.5\pm2.5$  & $-41.6\pm5.5$  & $20.1\pm5.5$  &$145.4\pm38.3$ & $920.9\pm148.2$ & $-2.1\pm0.4$ & $6.8$   & $0.8$  &  0.07   &  0.3  &   100 \\
9  & $-3.1\pm3.1$  & $-8.5\pm3.2$  & $4.8\pm2.9$    & $-65.3\pm2.9$ &$254.1\pm42.7$ & $937.9\pm187.3$&  $-1.3\pm0.2$ & $11.1$  & $7.5$  &  1.20   &  1.9  &   100\\
12 & $7.3\pm2.6$   & $21.4\pm2.9$  & $19.4\pm2.8$   & $38.1\pm2.8$  &$482.4\pm69.9$ & $735.7\pm117.4$&  $-1.2\pm0.2$ & $27.7$  & $16.2$  &  3.77   &  5.5  &   81.1\\
13 & $-2.0\pm2.5$  & $-5.6\pm2.7$  & $39.6\pm3.0$   & $-19.1\pm3.0$ &$185.7\pm35.8$ & $716.7\pm110.0$ & $2.8\pm0.5$  & $7.6$   & $3.6$  &  4.42   &  8.9  &   100 \\
14 & $-0.9\pm2.9$  & $-1.3\pm2.4$  & $-5.7\pm3.9$   & $-44.5\pm3.9$ &$200.2\pm50.9$ & $840.6\pm194.6$ & $-3.3\pm0.6$ & $7.4$   & $1.0$  &  5.86   &  7.4  &   100 \\
15 & $1.1\pm2.2$   & $-2.2\pm2.3$  & $-58.6\pm5.4$  & $8.1\pm5.4$   &$239.5\pm18.1$ & $656.7\pm117.5$&  $1.5\pm0.3$  & $11.2$  & $7.1$  &  15.98  &  21.7 &   100\\
16 & $-5.2\pm4.6$  & $1.1\pm5.1$   & $0.8\pm5.7$    & $-58.1\pm5.7$ &$293.3\pm74.6$ & $676.8\pm113.2$&  $1.1\pm0.1$  & $19.3$  & $2.4$  &  19.70  &  12.0 &   99.9   \\
17 & $12.5\pm3.2$  & $13.0\pm3.5$  & $25.2\pm2.5$   & $42.1\pm2.5$  &$346.8\pm30.9$ & $603.0\pm37.3$ &  $0.4\pm0.1$  & $10.3$  & $9.8$   &  20.01  &  10.9 &   100   \\
18 & $1.8\pm2.3$   & $-28.6\pm3.5$ & $7.2\pm3.1$    & $-32.2\pm3.1$ &$537.4\pm123.3$ & $651.1\pm134.8$& $2.7\pm0.5$  & $3.6$   & $0.5$ &   21.30 &  29.8 &   63.6  \\
19 & $-9.0\pm2.0$  & $-2.8\pm1.7$  & $-28.8\pm6.9$  & $8.7\pm6.9$   &$150.2\pm41.6$ & $644.5\pm188.0$ & $2.9\pm0.5$  & $3.6$   & $0.9$  &  23.69  &  34.6 &   100   \\
20 & $-0.4\pm2.4$  & $-1.0\pm3.1$  & $15.4\pm5.8$   & $-5.8\pm5.8$  &$255.7\pm69.0$ & $630.2\pm178.2$ & $4.6\pm0.8$  & $7.8$   & $1.9$  &  43.24  &  43.2 &   100   \\
\end{tabular}                                                  
\centering
\caption{\label{tab_pm}Comparison of our measured proper motions and those from the SDSS database as used by P14 (columns 2--5).
Galactic rest-frame velocities are in c.6\&7 based on our proper motions $\varv_\text{GRF,1}$ or SDSS proper motions $\varv_\text{GRF,2}$. 
The present height above the Galactic plane is denoted with $z$.
The quantities $r$ and $r_\text{min}$ give the average and minimum distance to the GC at the 3$\sigma$ level of the disk passage using new proper motions and model II of \citet{Irrgang2010}. 
The last three columns give the bound probabilities as listed by P14 (c.10), our results based on model II and the proper motions from SDSS as given in column 4\&5 (c.11) and based on our revised proper motions from column 2\&3 in the last column.}
\end{table*}

\section{Kinematics}
\label{sec_kinematics}
 
To obtain the dynamical properties of the stars, the orbits were traced back to the Galactic disk.
Three different Milky Way mass models, see models I, II, and III in \citet{Irrgang2013}, were used for this exercise to estimate possible systematic influences of the applied gravitational potential. 
Because the escape velocities of Model II of \citet{Irrgang2013} are very similar to those calculated by P14, we use this model for the comparison with the results of P14.

Varying the position and the velocity components within their respective errors by applying a Monte Carlo procedure with a depth of $10^6$, we determined the intersection area of the trajectories with the Galactic plane.
From these simulations we also derived the median Galactic rest-frame velocities at the present locations and their distributions.
These values are compared with the respective local escape velocity.
Due to our lower tangential velocity the stars show a much lower probability of being unbound to the Galaxy in all three applied gravitational potentials.
The bound probability is defined as the number of orbits not exceeding the local escape velocity with respect to the number of all calculated orbits.
In Table \ref{tab_pm} the bound probabilities listed by P14 is compared to the bound probabilities obtained by applying model II of \citet{Irrgang2013} using proper motions of SDSS, as used by P14, and of this work.
The bound probablities calculated using SDSS proper motions are similar to those obtained by P14, as expected.
For the Galactic models I (see Table \ref{tab_I}) and III (see Table \ref{tab_III}) the differences between P14 and our calculations using SDSS proper motions can be explained by the higher Galaxy masses of those models.

\begin{figure}[t]
\begin{center}
\includegraphics[scale=0.55]{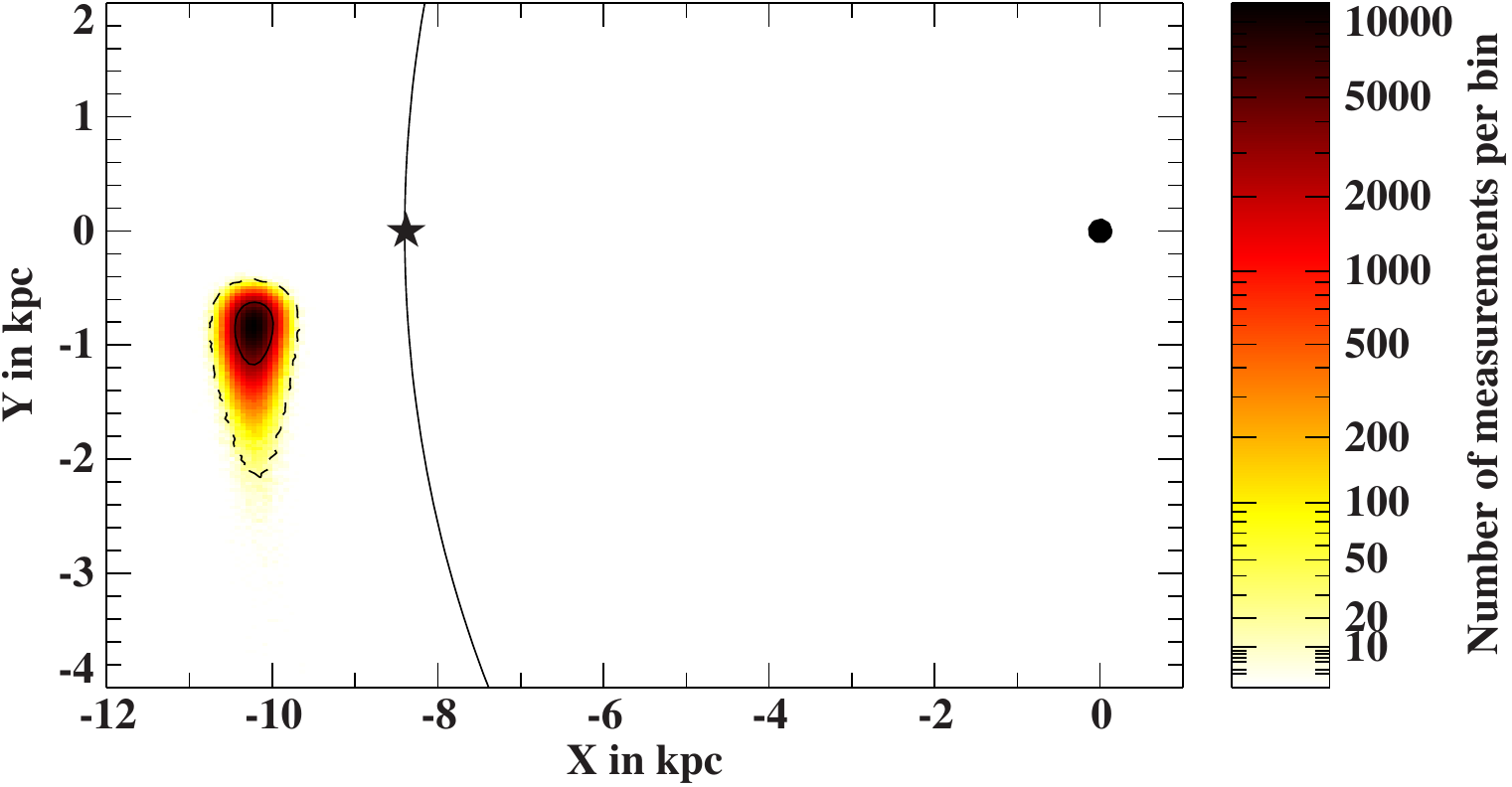}
\caption{\label{fig_18_origin}Disk passages (binned and color coded) of No. 17 obtained by a Monte Carlo simulation with 1 (solid) and 3 (dashed) $\sigma$ contours. The solid line indicates the solar orbit. The star marks the position of the Sun, the black dot that of the GC. }
\end{center}
\end{figure}

Using our revised proper motions, none of the 14 candidates is unbound irrespective of the Galactic potential model used. In fact, the probability to be unbound is less than 0.1\% for all but three candidates.
Star No.~18 has the highest probability of being unbound, but even in the lightest Galaxy model II this probability does not exceed 36\%.
We also calculate the distances to the GC at the time of the disk passage and the minimum distances in all three potentials (Table \ref{tab_pm}-\ref{tab_III}).
They do not depend on the choice of the potential except for Nos.~12, 16 and particularly No.~2 (the latter changes by almost a factor of 2), for which the location of disk crossing must be regarded as highly uncertain.
The GC is excluded at 3$\sigma$ level for all stars irrespective of the choice of potential. 
Among the programme stars the present position of star No.~17 is closest to the Galactic plane and, hence, its disk crossing location outside of the solar circle can be well constrained (see Fig. \ref{fig_18_origin}).

\section{Runaway, disk or halo stars?}
\label{sec_nature}

\begin{figure}[t]
\begin{center}
\includegraphics[scale=0.55]{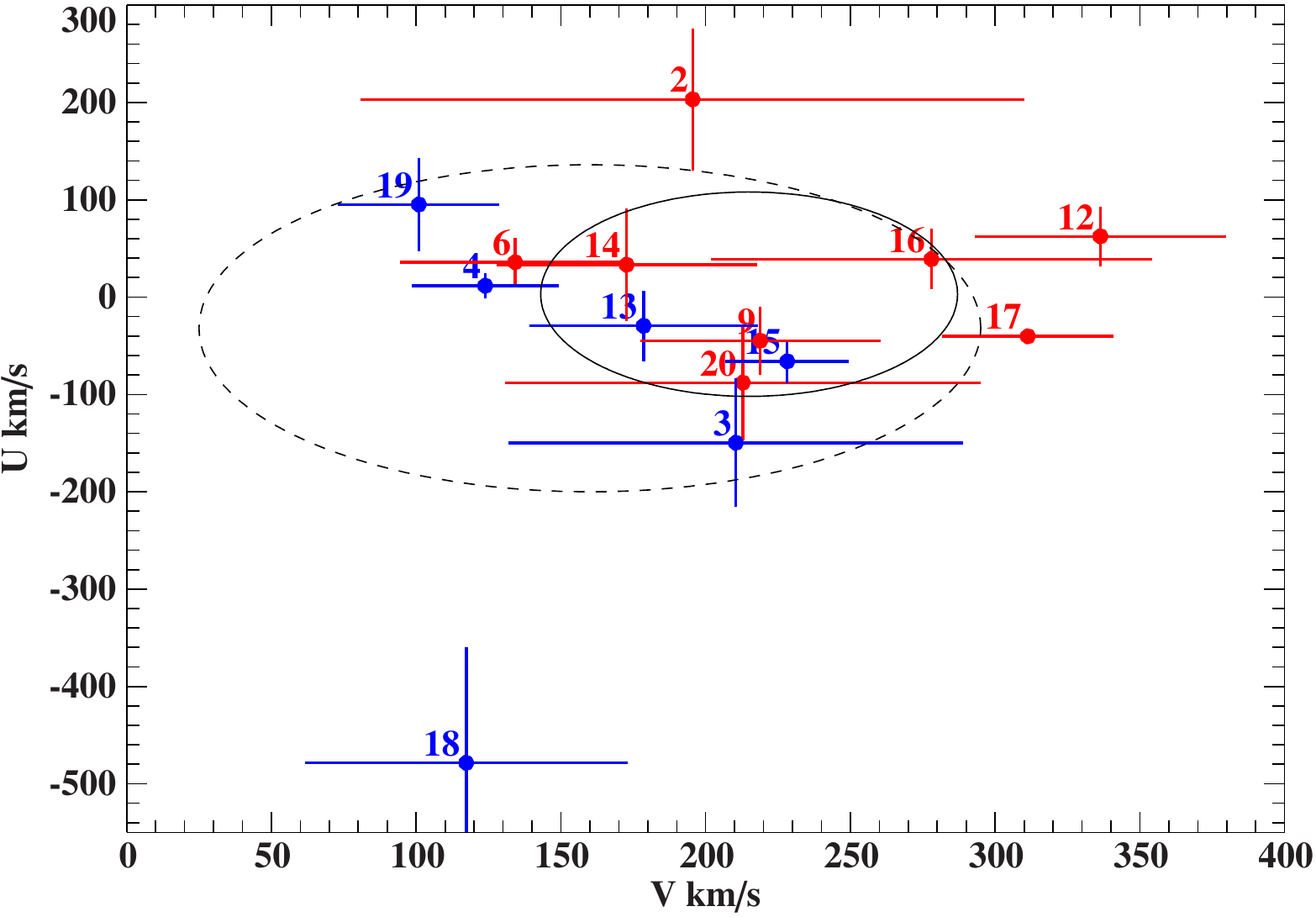}
\caption{\label{fig_UV}$U-V$-Diagram, numbered dots with error bars: stars examined in this work (red indicates stars with [Fe/H]$>-0.7$, blue [Fe/H]$<-0.7$), dashed line: $3\sigma$ contour of the thick disk, solid line: $3\sigma$ contour of the thin disk}
\end{center}
\end{figure}

\begin{figure}[t]
\begin{center}
\includegraphics[scale=0.78]{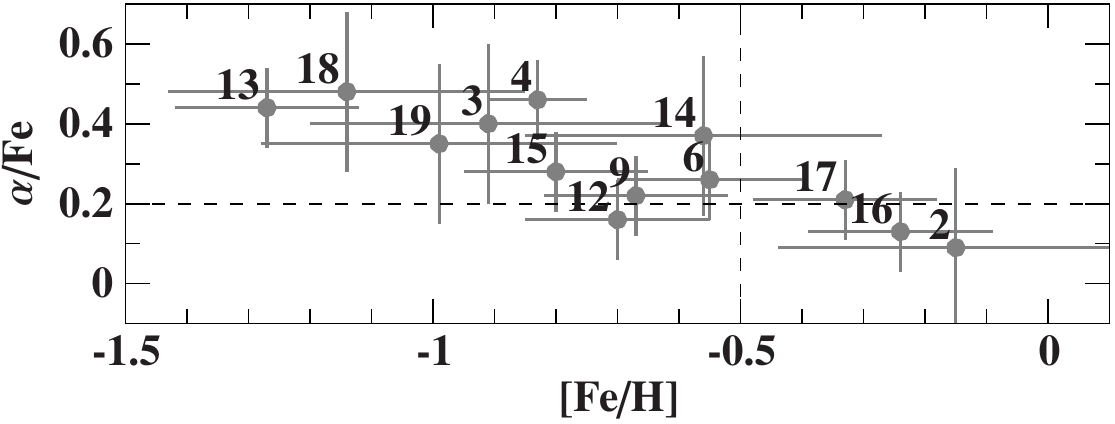}
\caption{\label{pic_alpha_fe}Position of the programme stars in the ([Fe/H], [$\alpha$/Fe]) diagram. The dashed lines are separating thick disk (top left) from thin disk stars (bottom right). Error bars were estimated from the S/N ratio of the spectra using the prescriptions of \citet{Allende2008} and \citet{Lee2011}.} 
\end{center}
\end{figure}

Since all 14 candidates are bound and none comes from anywhere near the GC, those stars have to be dismissed as HVS stars. 
P14 discussed whether they could be disk runaways or genuine halo stars. 
In Fig. \ref{fig_UV} the Galactic radial velocity component U is plotted against the rotational velocity component V and compared with contour lines denoting the limits for the thin and the thick disk, respectively by \citet{Pauli2006}.
However, kinematics cannot easily distinguish between thick disk and runaway origins.
Because most runaways are launched at modest velocities from the rotating disk, their kinematic and spatial distributions will naturally look quite similar to the thick disk \citep{Bromley2009}.
Population membership can be assigned also by chemical tagging, using [Fe/H] and [$\alpha$/Fe] as indicators.
According to \citet{Fuhrmann2011} thin disk stars have [Fe/H] $>$-0.5 and [$\alpha$/Fe]$<$0.2, whereas thick disk stars have [Fe/H] $<$-0.3 and [$\alpha$/Fe]$>$0.2 \citep[see Fig. 15 of][]{Fuhrmann2011}. 
However, some halo stars may have similar chemical abundances as thick disk stars. 
In Fig. \ref{pic_alpha_fe} we plot [$\alpha$/Fe] vs. [Fe/H] for the 14 stars and conclude that most stars have chemical characterics of the thick disk, whereas Nos. 2, 16, and 17 are more likely to be thin disk stars. 
The chemical composition of No. 18 strengthens the conclusion that it is a halo star.

Dynamical ejection in dense star clusters or binary supernova ejection may result in disk runaway stars. 
\citet{Kenyon2014} simulated the Galactic populations of (i) HVS stars ejected from the GC via the SMBH slingshot mechanism, (ii) runaway stars from the disk ejected at distances of 3 to 30 kpc from the GC by either the binary supernova mechanism or dynamcial ejection from clusters. 
They compare their predictions to the observed properties of B-type HVS \citep{Brown2014} and B-type runaway stars \citep{Silva2011} as well as to the P14 sample of G- and K-type HVS candidates.
\citet{Kenyon2014} conclude that the runaway sample of B-type stars can be matched by their simulation for binary supernova ejection or dynamical ejection, but not by the SMBH slingshot. 
The HVS sample of B-type stars \citep{Brown2014}, on the other hand, is well reproduced by the prediction from the SMBH slingshot, but the data can not be explained by runaway models.

\begin{figure}[t]
\begin{center}
\includegraphics[scale=0.55]{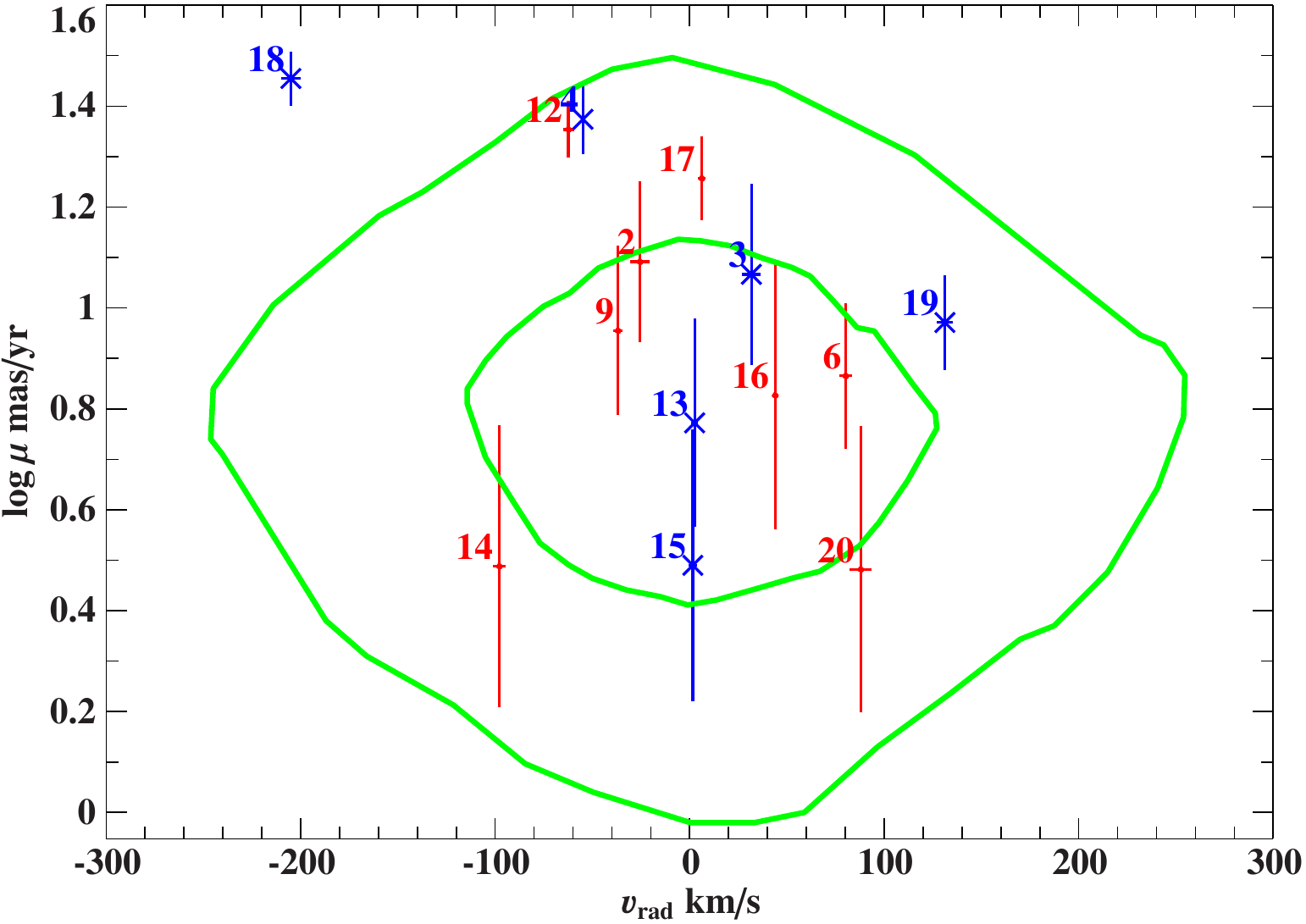}
\caption{\label{fig_Kenyon} Comparison of the HVS candidates from P14 with proper motions derived by us (red indicates stars with [Fe/H]$>-0.7$, blue crosses [Fe/H]$<-0.7$) in the proper motion vs. radial velocity plane with predictions of simulations for a sample of $1M_\odot$ runaways (green curves) generated by the binary supernova mechanism \citep{Kenyon2014}. The inner contour includes $50\%$ of the stars in the distance-limited sample, while the outer one includes $90\%$ \citep[adapted from Fig. 25 of][]{Kenyon2014}.}
\end{center}
\end{figure}

However, SMBH slingshot 
models fail to account for the observations of the P14 SEGUE HVS candidates of spectral type G and K.
In Fig. \ref{fig_Kenyon} we compare the radial velocities and our new proper motions of the 14 P14 stars analysed here to the predictions from simulations for 1 M$_\odot$ runaway stars generated by the binary supernova mechanism.
The distribution of stars match the prediction of \citet{Kenyon2014} very well. 
However, the stars' chemical composition is inconsistent with the disk runaway scenario except possibly for the three most metal-rich ones.
Such stars could be ejected as the surviving donor of an exploding white dwarf in a SN Ia event as is discussed for Tycho's supernova \citep{Ruiz2004}.

\section{Summary and conclusions}
\label{sec_conclusions}

We present a revised analysis of the 20 candidate HVSs introduced by P14. 
Our work was motivated by the warning of P14 that the proper motions they extracted from the SDSS data base were abnormally high and therefore the stars have to be considered HVS candidates only until their proper motions are confirmed.  
Therefore, we independently determined proper motions from all astrometric images using grids of distant galaxies as frames of reference for 14 of these candidates. 
We repeated the analysis of P14 and confirmed the radial velocites, atmospheric parameters and distances. 
However, all measured proper motions, except for two stars, are significantly lower than those used by P14. 
We carried out a kinematic analysis of the stars in three different Milky Way mass models and confirmed that no star originated from the GC.
All stars turn out to be bound to the Galaxy and therefore have to be dismissed as HVS.
We considered the possibilities that the stars are either old stars of the halo or thick disk as suggested by their metallicities and $\alpha$ element enhancements from the (U,V) diagram or disk-run away stars. 
The stars' kinematics as well as their chemical composition indicate that they mostly belong to the thick disk population, accordingly No.~18 could belong to the halo. 
However, the disk runaway option would also be consistent with model predictions by \citet{Kenyon2014} for all stars (except No. 18). 
This, however, is at variance with the stars' chemical composition except for the three most metal rich ones, which could possibly be runaway stars from the thin disk.

\begin{acknowledgements}
E.Z. acknowledges funding by the German Science foundation (DFG) through grant HE1356/45-2. B.T.G. was supported by ERC Grant Agreement n. 320964 (WDTracer)
\end{acknowledgements}

\bibliography{aabib}

\begin{thebibliography}{23}
\expandafter\ifx\csname natexlab\endcsname\relax\def\natexlab#1{#1}\fi

\bibitem[{{Allende Prieto} {et~al.}(2008){Allende Prieto}, {Sivarani}, {Beers},
  {Lee}, {Koesterke}, {Shetrone}, {Sneden}, {Lambert}, {Wilhelm}, {Rockosi},
  {Lai}, {Yanny}, {Ivans}, {Johnson}, {Aoki}, {Bailer-Jones}, \& {Re
  Fiorentin}}]{Allende2008}
{Allende Prieto}, C., {Sivarani}, T., {Beers}, T.~C., {et~al.} 2008, \aj, 136,
  2070

\bibitem[{{Blaauw}(1961)}]{Blaauw1961}
{Blaauw}, A. 1961, \bain, 15, 265

\bibitem[{{Bromley} {et~al.}(2009){Bromley}, {Kenyon}, {Brown}, \&
  {Geller}}]{Bromley2009}
{Bromley}, B.~C., {Kenyon}, S.~J., {Brown}, W.~R., \& {Geller}, M.~J. 2009,
  \apj, 706, 925

\bibitem[{{Brown} {et~al.}(2014){Brown}, {Geller}, \& {Kenyon}}]{Brown2014}
{Brown}, W.~R., {Geller}, M.~J., \& {Kenyon}, S.~J. 2014, \apj, 787, 89

\bibitem[{{Brown} {et~al.}(2005){Brown}, {Geller}, {Kenyon}, \&
  {Kurtz}}]{Brown2005}
{Brown}, W.~R., {Geller}, M.~J., {Kenyon}, S.~J., \& {Kurtz}, M.~J. 2005,
  \apjl, 622, L33

\bibitem[{{Castelli}(1999)}]{Castelli1999}
{Castelli}, F. 1999, \aap, 346, 564

\bibitem[{{Edelmann} {et~al.}(2005){Edelmann}, {Napiwotzki}, {Heber},
  {Christlieb}, \& {Reimers}}]{Edelmann2005}
{Edelmann}, H., {Napiwotzki}, R., {Heber}, U., {Christlieb}, N., \& {Reimers},
  D. 2005, \apjl, 634, L181

\bibitem[{{Fuhrmann}(2011)}]{Fuhrmann2011}
{Fuhrmann}, K. 2011, \mnras, 414, 2893

\bibitem[{{Hawkins} {et~al.}(2015){Hawkins}, {Kordopatis}, {Gilmore},
  {Masseron}, {Wyse}, {Ruchti}, {Bienaym{\'e}}, {Bland-Hawthorn}, {Boeche},
  {Freeman}, {Gibson}, {Grebel}, {Helmi}, {Kunder}, {Munari}, {Navarro},
  {Parker}, {Reid}, {Scholz}, {Seabroke}, {Siebert}, {Steinmetz}, {Watson}, \&
  {Zwitter}}]{Hawkins2014}
{Hawkins}, K., {Kordopatis}, G., {Gilmore}, G., {et~al.} 2015, \mnras, 447,
  2046

\bibitem[{{Hills}(1988)}]{Hills1988}
{Hills}, J.~G. 1988, \nat, 331, 687

\bibitem[{{Hirsch} {et~al.}(2005){Hirsch}, {Heber}, {O'Toole}, \&
  {Bresolin}}]{Hirsch2005}
{Hirsch}, H.~A., {Heber}, U., {O'Toole}, S.~J., \& {Bresolin}, F. 2005, \aap,
  444, L61

\bibitem[{{Irrgang} {et~al.}(2010){Irrgang}, {Przybilla}, {Heber}, {Nieva}, \&
  {Schuh}}]{Irrgang2010}
{Irrgang}, A., {Przybilla}, N., {Heber}, U., {Nieva}, M.~F., \& {Schuh}, S.
  2010, 711, 138

\bibitem[{{Irrgang} {et~al.}(2013){Irrgang}, {Wilcox}, {Tucker}, \&
  {Schiefelbein}}]{Irrgang2013}
{Irrgang}, A., {Wilcox}, B., {Tucker}, E., \& {Schiefelbein}, L. 2013, \aap,
  549, A137

\bibitem[{{Kenyon} {et~al.}(2014){Kenyon}, {Bromley}, {Brown}, \&
  {Geller}}]{Kenyon2014}
{Kenyon}, S.~J., {Bromley}, B.~C., {Brown}, W.~R., \& {Geller}, M.~J. 2014,
  ArXiv e-prints

\bibitem[{{Lee} {et~al.}(2011){Lee}, {Beers}, {Allende Prieto}, {Lai},
  {Rockosi}, {Morrison}, {Johnson}, {An}, {Sivarani}, \& {Yanny}}]{Lee2011}
{Lee}, Y.~S., {Beers}, T.~C., {Allende Prieto}, C., {et~al.} 2011, \aj, 141, 90

\bibitem[{{Munari} {et~al.}(2005){Munari}, {Sordo}, {Castelli}, \&
  {Zwitter}}]{Munari2005}
{Munari}, U., {Sordo}, R., {Castelli}, F., \& {Zwitter}, T. 2005, \aap, 442,
  1127

\bibitem[{{Palladino} {et~al.}(2014){Palladino}, {Schlesinger},
  {Holley-Bockelmann}, {Allende Prieto}, {Beers}, {Lee}, \&
  {Schneider}}]{Palladino2014}
{Palladino}, L.~E., {Schlesinger}, K.~J., {Holley-Bockelmann}, K., {et~al.}
  2014, \apj, 780, 7

\bibitem[{{Pauli} {et~al.}(2006){Pauli}, {Napiwotzki}, {Heber}, {Altmann}, \&
  {Odenkirchen}}]{Pauli2006}
{Pauli}, E.-M., {Napiwotzki}, R., {Heber}, U., {Altmann}, M., \& {Odenkirchen},
  M. 2006, \aap, 447, 173

\bibitem[{{Ruiz-Lapuente} {et~al.}(2004){Ruiz-Lapuente}, {Comeron},
  {M{\'e}ndez}, {Canal}, {Smartt}, {Filippenko}, {Kurucz}, {Chornock}, {Foley},
  {Stanishev}, \& {Ibata}}]{Ruiz2004}
{Ruiz-Lapuente}, P., {Comeron}, F., {M{\'e}ndez}, J., {et~al.} 2004, \nat, 431,
  1069

\bibitem[{{Silva} \& {Napiwotzki}(2011)}]{Silva2011}
{Silva}, M.~D.~V. \& {Napiwotzki}, R. 2011, \mnras, 411, 2596

\bibitem[{{Tauris}(2015)}]{Tauris2014}
{Tauris}, T.~M. 2015, \mnras, 448, L6

\bibitem[{{Tillich} {et~al.}(2011){Tillich}, {Heber}, {Geier}, {Hirsch},
  {Maxted}, {G{\"a}nsicke}, {Marsh}, {Napiwotzki}, {{\O}stensen}, \&
  {Scholz}}]{Tillich2011}
{Tillich}, A., {Heber}, U., {Geier}, S., {et~al.} 2011, \aap, 527, A137

\bibitem[{{Zhong} {et~al.}(2014){Zhong}, {Chen}, {Liu}, {de Grijs}, {Hou},
  {Shen}, {Shao}, {Li}, {Luo}, {Shi}, {Zhang}, {Yang}, {Deng}, {Jin}, {Zhang},
  {Hou}, \& {Zhang}}]{Zhong2014}
{Zhong}, J., {Chen}, L., {Liu}, C., {et~al.} 2014, \apjl, 789, L2

\end{thebibliography}
\bibliographystyle{aa}

\Online

\begin{table}
\scriptsize
\setlength{\tabcolsep}{1mm}
\begin{tabular}{l|cc|cc|ccc}
 & & & & & \multicolumn{3}{|c}{bound probabilities (\%)}\\
P14 & $r$ & $r_\text{min}$ & $\varv_\text{GRF,1}$ & $\varv_\text{GRF,2}$ & P14 & \multicolumn{2}{c}{model I}\\
No. & (kpc) & (kpc) & (km/s) & (km/s) & \multicolumn{2}{c}{PM: SDSS} & this work \\
\hline\hline
2  & $50.3$ & $3.8$ & $424.5\pm100.5$ & $636.4\pm86.7$  & 7.43   &  23.4 &    92.3  \\
3   & $8.5$ & $1.3$ & $304.6\pm75.8$ & $647.2\pm96.7$   & 34.88  &  49.3 &   100    \\
4   & $6.7$ & $3.9$ & $128.1\pm23.2$ & $1305.8\pm213.4$ & 0.00   &  0.0  &   100    \\
6   & $6.8$ & $0.9$ & $146.7\pm38.4$ & $922.6\pm147.9$  & 0.07   &  0.8  &    100   \\
9  & $11.1$ & $7.5$ & $255.4\pm42.6$ & $937.5\pm187.1$  & 1.20   &  3.3  &    100   \\
12 & $24.2$ & $16.0$ & $483.5\pm69.9$ & $736.4\pm117.4$ & 3.77   &  11.0 &    91.7  \\
13  & $7.6$ & $3.6$ & $187.2\pm35.8$ & $717.7\pm109.8$ & 4.42   &  16.9 &    100   \\
14  & $7.4$ & $1.0$ & $201.5\pm50.9$ & $839.7\pm194.4$ & 5.86   &  11.3 &    100   \\
15 & $11.1$ & $7.1$ & $241.1\pm18.1$ & $657.0\pm117.3$ & 15.98  &  33.8 &    100   \\
16 & $18.6$ & $2.5$ & $294.5\pm74.4$ & $675.9\pm113.3$ & 19.70  &  22.2 &    100   \\
17 & $10.3$ & $9.8$ & $348.2\pm30.9$ & $604.4\pm37.3$  & 20.01  &  44.4 &    100   \\
18  & $3.5$ & $0.5$ & $537.1\pm123.5$ & $651.1\pm134.9$&  21.30 &  41.4 &    74.6  \\
19  & $3.7$ & $1.0$ & $150.7\pm41.3$ & $644.6\pm187.9$ & 23.69  &  43.4 &    100   \\
20  & $7.9$ & $2.0$ & $257.0\pm69.0$ & $631.7\pm178.0$ & 43.24  &  52.0 &    100   \\
\end{tabular}      
\centering
\caption{\label{tab_I}The quantities $r$ and $r_\text{min}$ give the average and minimum distance to the GC at the 3$\sigma$ level of the disk passage using new proper motions and model I of \citet{Irrgang2010}.  Galactic rest-frame velocities $\varv_\text{GRF}$ based on our proper motions (1) or SDSS proper motions (2). The last three columns give the bound probabilities as listed by P14 (c.6), our results based on model I and the proper motions from SDSS as given in column 4\&5 (Table \ref{tab_pm}) (c.7) and based on our revised proper motions from column 2\&3 (Table \ref{tab_pm}) in the last column.}
\end{table}

\begin{table}
\scriptsize
\setlength{\tabcolsep}{1mm}
\begin{tabular}{l|cc|cc|ccc}
 & & & & & \multicolumn{3}{|c}{bound probabilities}\\
P14 & $r$ & $r_\text{min}$ & $\varv_\text{GRF,1}$ & $\varv_\text{GRF,2}$ & P14 & \multicolumn{2}{c}{model III}\\
No. & (kpc) & (kpc) & (km/s) & (km/s) & \multicolumn{2}{c}{PM: SDSS} & this work \\
\hline\hline
2  & $34.1$ & $4.0$ & $423.3\pm100.5$ & $634.4\pm86.7$  & 7.43   &  94.4 &   99.9  \\
3   & $7.9$ & $1.1$ & $303.2\pm75.8$ & $645.3\pm96.9$   & 34.88  &  96.3 &   100   \\
4   & $6.7$ & $3.9$ & $126.5\pm23.1$ & $1304.1\pm213.8$ & 0.00   &  1.0  &   100   \\
6   & $6.8$ & $0.9$ & $144.7\pm38.4$ & $920.5\pm147.8$  & 0.07   &  19.8 &   100   \\
9  & $11.1$ & $7.5$ & $253.5\pm42.7$ & $937.8\pm187.4$  & 1.20   &  22.6 &   100   \\
12 & $18.9$ & $14.2$ & $481.8\pm69.9$ & $735.1\pm117.6$ & 3.77   &  68.4 &   100   \\
13  & $7.6$ & $3.7$ & $185.1\pm35.8$ & $716.6\pm110.0$ & 4.42   &  79.4 &   100   \\
14  & $7.3$ & $1.1$ & $199.6\pm51.0$ & $841.0\pm194.5$ & 5.86   &  43.3 &   100   \\
15 & $11.2$ & $7.2$ & $239.5\pm18.1$ & $656.5\pm117.5$ & 15.98  &  88.3 &   100   \\
16 & $16.3$ & $2.5$ & $292.5\pm74.5$ & $677.7\pm113.3$ & 19.70  &  83.5 &   100   \\
17 & $10.3$ & $9.8$ & $346.2\pm30.9$ & $602.5\pm37.3$  & 20.01  &  100.0 &   100   \\
18  & $3.6$ & $0.5$ & $537.4\pm123.3$ & $651.1\pm134.8$&  21.30 &  87.7 &   97.9  \\
19  & $3.7$ & $1.0$ & $149.8\pm41.7$ & $644.1\pm187.9$ & 23.69  &  80.0 &   100   \\
20  & $8.0$ & $2.1$ & $255.1\pm68.8$ & $630.2\pm178.4$ & 43.24  &  85.4 &   100   \\
\end{tabular}         
\centering
\caption{\label{tab_III}The quantities $r$ and $r_\text{min}$ give the average and minimum distance to the GC at the 3$\sigma$ level of the disk passage using new proper motions and model III of \citet{Irrgang2010}.  Galactic rest-frame velocities $\varv_\text{GRF}$ based on our proper motions (1) or SDSS proper motions (2). The last three columns give the bound probabilities as listed by P14 (c.6), our results based on model III and the proper motions from SDSS as given in column 4\&5 (Table \ref{tab_pm}) (c.7) and based on our revised proper motions from column 2\&3 (Table \ref{tab_pm}) in the last column.}
\end{table}

\end{document}